\pdfoutput=1
\documentclass[usenatbib]{mn2e}
\usepackage{apjfonts}
\usepackage{amssymb}
\usepackage{amsmath}
\usepackage{ctable}
\usepackage{fixltx2e} %% forces figure order to remain fixed (otherwise figure*'s can move out-of-order)

\newcommand{\be}{\begin{equation}}
\newcommand{\ee}{\end{equation}}

\newcommand{\msun}{M_{\sun}}

\newcommand{\degree}{^{\circ}}
\newcommand{\paperone}{Paper {\small I}}
\newcommand{\papertwo}{Paper {\small II}}
\newcommand{\paperthree}{Paper {\small III}}

\newcommand\plotonesize[2]
 {\centering \leavevmode \includegraphics[width={#2\columnwidth}]{#1}}

\newcommand{\acknowledgments}{\begin{small}\section*{Acknowledgments}\end{small}}
\newcommand\altaffilmark[1]{$^{#1}$}
\newcommand\altaffiltext[1]{$^{#1}$}
\title[Accretion Does Not Drive Turbulence]{Accretion Does Not Drive the Turbulence in Galactic Disks\vspace{-0.5cm}}
\author[Hopkins et al.]{
\parbox[t]{\textwidth}{ 
Philip F.~Hopkins\thanks{E-mail:phopkins@caltech.edu}\altaffilmark{1,2},
Du\v{s}an Kere\v{s}\altaffilmark{3}, 
\&\ Norman Murray\altaffilmark{4,5} 
} 
\vspace*{6pt} \\
\altaffiltext{1}{TAPIR, Mailcode 350-17, California Institute of Technology, Pasadena, CA 91125, USA} \\
\altaffiltext{2}{Department of Astronomy and Theoretical Astrophysics Center, University of California Berkeley, Berkeley, CA 94720} \\
\altaffiltext{3}{Department of Physics, Center for Astrophysics and Space Science, University of California at San Diego, 9500 Gilman Drive, La Jolla, CA 92093} \\ 
  \altaffiltext{4}{Canadian Institute for Theoretical Astrophysics, 
60 St.\ George Street, University of Toronto, ON M5S 3H8, Canada} \\
\altaffiltext{5}{Canada Research Chair in Astrophysics\vspace{-0.7cm}} 
}

\date{Submitted to MNRAS, December, 2012\vspace{-0.6cm}}
\begin{document}
\maketitle
\label{firstpage}

\begin{abstract}

Rapid accretion of cold intergalactic gas plays a crucial role in getting gas into galaxies. It has been suggested that this gas accretion proceeds along  narrow streams that might also directly drive the turbulence in galactic gas, dynamical disturbances, and bulge formation. In cosmological simulations, however, it is impossible to isolate and hence disentangle the effect of cold stream accretion from internal instabilities and mergers. Moreover, in most current cosmological simulations, the phase structure and turbulence in the interstellar medium (ISM) arising from stellar feedback are treated in an approximate (sub-grid) manner, so that the feedback cannot generate turbulence in the ISM. In this paper we therefore test the effects of cold streams in extremely high-resolution simulations of otherwise isolated galaxy disks using detailed models for star formation and stellar feedback; we then include or exclude mock cold flows falling onto the galaxies, with mass accretion rates, velocities and flow geometry set to maximize their effect on the gaseous disk. We find: (1) Turbulent velocity dispersions in gas disks are identical with or without the presence of the cold flow; the energy injected by the flow is efficiently dissipated where it meets the disk. (2) In runs without stellar feedback, the presence of a cold flow has essentially no effect on runaway fragmentation (local collapse), resulting in star formation rates (SFRs) that are an order-of-magnitude too large. (3) Model disks in runs with both explicit feedback and cold flows have higher SFRs, but only insofar as they have more gas. (4) Because the flows are extended, relative to the size of the disk, they do not trigger strong resonant responses and so induce weak gross morphological perturbation (bulge formation via instabilities/fragmentation is not accelerated). (5) However, flows can thicken the disk by direct contribution of out-of-plane or mis-aligned star-forming streams/filaments. We conclude that while inflows are critical over cosmological timescales to determine the supply and angular momentum of gas disks, they have  weak {\em instantaneous} dynamical effects on galaxies. 

\end{abstract}

\begin{keywords}
galaxies: formation --- galaxies: evolution --- galaxies: active --- 
star formation: general --- cosmology: theory
\vspace{-1.0cm}
\end{keywords}

\vspace{-1.1cm}
\section{Introduction}
\label{sec:intro}

%Silk 1977 was unaware of CDM (the paper assumes baryons only)

Since some of the earliest models of galaxy formation in a cosmological setting, it has been known that in low-mass halos ($\lesssim 10^{11-12}\,\msun$), the gas cooling time is shorter than the dynamical time even if gas is shock heated to virial temperature before accretion. In this case shock heating during virialization will not provide sufficient pressure support for the infalling gas and gas could reach galaxies in near free-fall (\citealt{silk:1977.galaxy.cooling.fragmentation,rees:1977.tcool.tdyn.vs.mhalo,binney:1977.weak.vir.shocks,whiterees78}; for a more detailed spherical shock stability analysis with a modern $\Lambda$CDM cosmology, see \citealt{birnboim:mquench}). In recent years, numerical hydrodynamic simulations have considerably improved our understanding of this regime which is part of a more general process termed ``cold accretion'' or ``cold flows.'' \citet{keres:hot.halos} argued that in lower mass systems, gas can reach galaxies in free-falling filaments which are cold ($\sim10^{4}-10^{5}\,$K) by virtue of their large densities preventing an accretion shock; at high redshifts ($z\gtrsim2$), such streams may even penetrate more massive halos (where the spherically-averaged $t_{\rm cool}\gg t_{\rm dyn}$) because of the highly non-spherical nature of accretion. In this early work the gas accreted by galaxies in massive halos was shown to largely come from the ``hot mode accretion'', i.e. most of the infalling filamentary gas heats before reaching galaxies. A large body of subsequent theoretical work has developed on this topic \citep[see e.g.][and references therein]{dekelbirnboim:mquench,dekel:cold.streams,oppenheimer:recycled.wind.accretion,ocvirk:2008.cold.accretion,keres:cooling.revised,brooks:2009.coldflow.disk.assembly}. At the same time, tantalizing observational evidence for the existence of infalling cool gas is emerging \citep{ribaudo:2011.cold.flow.evidence,kacprzak:2012.accretion.flow.evidence}, although systematic detection of large gas infall rates in high-redshift halos is still lacking \citep{sancisi:2008.col.accretion,steidel:2010.outflow.kinematics}. 
%Above I removed the maller&bullock references thay do not study cosmological infall

%There is little theoretical debate that this occurs on some level (although observations remain elusive; see \citealt{sancisi:2008.col.accretion} and references therein), and that it is a primary means by which many star-forming galaxies -- especially low-mass and/or high-redshift systems -- obtain their gas (in the sense that it transports the gas, as opposed to the gas being carried into the galaxy entirely within bound galaxy merger ``units,'' for example). 

The exact nature of the accretion process in massive ($\sim 10^{12}\msun$) halos is still a subject of a debate, owing to the sensitivity of predictions to details of the treatment of gas and ``feedback'' physics. All simulations agree that most of the gas in massive halos heats to high temperatures. The debate is mostly about the detailed nature of the infalling gas in the vicinity of galaxies. At $z\sim 2$ cold streams of gas can still bring material deeply into the halos, however most of this gas may be heated before it is accreted to the halo center \citep{keres:hot.halos, ocvirk:2008.cold.accretion}. Subsequent work \citep{keres:cooling.revised} using different hydrodynamic codes found that only a relatively small fraction (typically $\sim10-20\%$ of the total halo infall at $z\sim2$, see \citealt{faucher-giguere:2011.halo.inflow.properties}) survives all the way to the central galaxy. Cooling of the hot halo in these simulations was not efficient, so this small fraction of surviving cold gas was found to dominate the gas supply to the halo center. Using another technique, \citet{dekel:cold.streams} suggested that at $z\sim 2-3$ rapid infall continues unimpeded via narrow filaments all the way to the halo center even for massive halos.

Recent simulations using the adaptive mesh refinement and the moving mesh code AREPO (which resolves some numerical inaccuracies previous works using the ``classical SPH'' formulation of GADGET; see \citealt{springel:arepo}) find that (absent galactic outflows) the infalling gas is indeed filamentary within the virial radius but may be heated before accretion onto the galaxy \citep{nelson:2013.arepo.coldflow.structure}. In these runs, at $z\sim2$ a large fraction of gas in a $\sim10^{12}\msun$ halo flows as cool gas down to $\sim 0.3\,R_{\rm vir}$; the infalling gas then shock-heats, but remains in a coherent structure before it cools again rapidly close to the galaxy \citep[see][]{keres:2011.arepo.gadget.disk.angmom,nelson:2013.arepo.coldflow.structure}. However, the gas heating removes some kinetic energy of the infall and forms structures whose coherence lengths exceed the size of the galactic disk ($\sim10\,$kpc). Further complexity comes from the halo mass dependence; in lower-mass halos the width of filaments is typically large compared to the galaxy (even when filaments are cold). Thus, the simplest picture of narrow streams of cold gas impacting the galactic disk should only be considered approximate \citep[see discussions in][]{katz:2003.how.gal.get.gas,keres:cooling.revised,nelson:2013.arepo.coldflow.structure}.

%However gas heating removes some of the kinetic energy of the infall and forms structures whose coherence lengths exceed the size of galactic disk. ($R_{\rm disk}\sim 10\kpc$). Thus the picture of narrow streams of cold gas impacting the disk may need to be modified for $10^{12}M_\odot$ halos at least at z$\sim$2, redshift regime where velocity dispersion of the gas in the disk is observed to be very high \citep{genzel:highz.rapid.secular}. Further complexity comes from the halo mass dependence: in lower mass halos and at lower redshift the width of the filaments is typically large compared to the galaxy, even when filaments are cold, which means that the infall is less likely to be seen as narrow streams of gas (see discussions in \citet{katz:2003.how.gal.get.gas,keres:cooling.revised,nelson:2013.arepo.coldflow.structure}). 

In any case, the consequences of narrow filamentary inflows -- assuming they exist -- are controversial. For example, it is well-known that in a gaseous disk where cooling is efficient, turbulent velocity dispersions should dissipate in a crossing time, so something must continuously maintain the turbulence in galaxies. Although this is often attributed to a combination of stellar feedback and {local} gravitational instabilities, some authors have argued that the observed turbulent velocity dispersions -- especially in high-redshift galaxies (when the galaxies are highly gas-rich, as distinct from systems like the Milky Way) -- may be driven in part by the accretion energy \citep[for discussion of this possibility, see e.g.][]{elmegreen:2010.accretion.to.turb,klessen:2010.accretion.turb,krumholz:2010.instab.turb.in.disks,cacciato:2011.analytic.disk.instab.cosmo.evol}. By extension (if the turbulence can be both driven and efficiently isotropized) this could argue for turbulence driving gas disk thicknesses/scale-heights. In some versions of this argument, maintaining the large-scale turbulence would in turn be sufficient to explain the low star formation efficiencies of galaxies (i.e.\ the \citealt{kennicutt98} relation). Others have argued that the flows directly trigger or enhance secular instabilities in sufficiently gas-rich disks, driving bulge formation \citep[though this may depend on the flow structure, as we discuss in \S~\ref{sec:discussion}; see][]{dekel:cold.streams,elmegreen:2010.accretion.to.turb}.\footnote{We should caution that ``accretion driven turbulence'' can have many different meanings, and is used differently in some of these examples. Much of the focus on these papers is on the role of accretion {\em through} a galactic disk, powered itself via gravitational instabilities, and its relation to turbulence, or accretion {\em within} the disk onto individual molecular clouds. That is not the focus of this paper. Rather, we investigate the very specific question of the role of accretion {\em onto} the galaxy.}

If it were true that accretion {\em onto} the galaxy drives turbulence in an {\em instantaneous} sense, this this would imply a radical departure from one of the fundamental assumptions in most analytic or semi-analytic galaxy formation models (indeed including many of the models discussed above). In such models, accretion is the ``fuel supply'' which determines the gas mass, angular momentum content, radius, etc.\ of the galaxy disk; these properties in turn determine things like the star formation rate and local stability properties, hence ultimately the disk structure and feedback properties. But the inflowing gas, excluding discrete merger events, does not {\em directly} alter the structural properties of the disk.  If the inflow powers the turbulence in the disk, for otherwise {\em identical} properties in that disk, then the instantaneous properties of the ``accretion flow'' will significantly change the gas velocity dispersions within the disk, and hence all the disk properties that depend on the velocity dispersion (e.g., the star formation rate and stability).

It is critical to assess this claim in numerical simulations, which can simultaneously follow the relevant physics: inflow, gas cooling, star formation, stellar feedback and the formation of ISM phase structure. Such simulations can follow the complex, non-linear dynamics of turbulence, torques, and energy and angular momentum exchange in gas+stellar+dark matter disk galaxies. However, in traditional cosmological simulations -- which can self-consistently follow the formation of inflows and their properties -- one cannot simply ``remove'' the accretion flows, nor vary their properties systematically while keeping the galaxy properties fixed. In other words, cosmological simulations are not controlled experiments as far as accretion flows are concerned. 

In addition, achieving the resolution needed to resolve the internal structural properties of the gas disk (resolving the vertical turbulent cascade, for example; see \citealt{federrath:2011.gravity.driven.turb}) is very challenging in a cosmological simulation, with simulation volumes measured in tens of megaparsecs or larger. Even in smaller volume ``zoom-in'' simulation, the timescales which must be simulated (a Hubble time) severely limits the attainable force resolution. Finally, it is well-known that the gas properties are strongly affected by feedback from massive stars, so including explicit models for these feedback properties is critical. 

In this paper, we therefore use a suite of high resolution hydrodynamic simulations of isolated disk galaxies (with properties chosen to match observations) and explicit stellar feedback models, including various simplified ``cold flow'' models of infalling gas, to investigate these questions. In order to study the influence of infalling gas on the gas dynamics in galaxies, we will consider the most optimistic assumptions (ignoring some of the subtleties above). We will assume that narrow filaments of cool gas do manage to survive and plunge into the disk of the galaxy. Furthermore, we assume that infall rates from cold mode accretion onto galaxies represent the full infall rate onto halos (clearly an upper limit). Our purpose is to see if such filamentary accretion can drive high velocity dispersion and scale-heights of galaxies, and to assess the relative importance of this process when compared to star formation-driven feedback.

\begin{figure}
    \centering
    \plotonesize{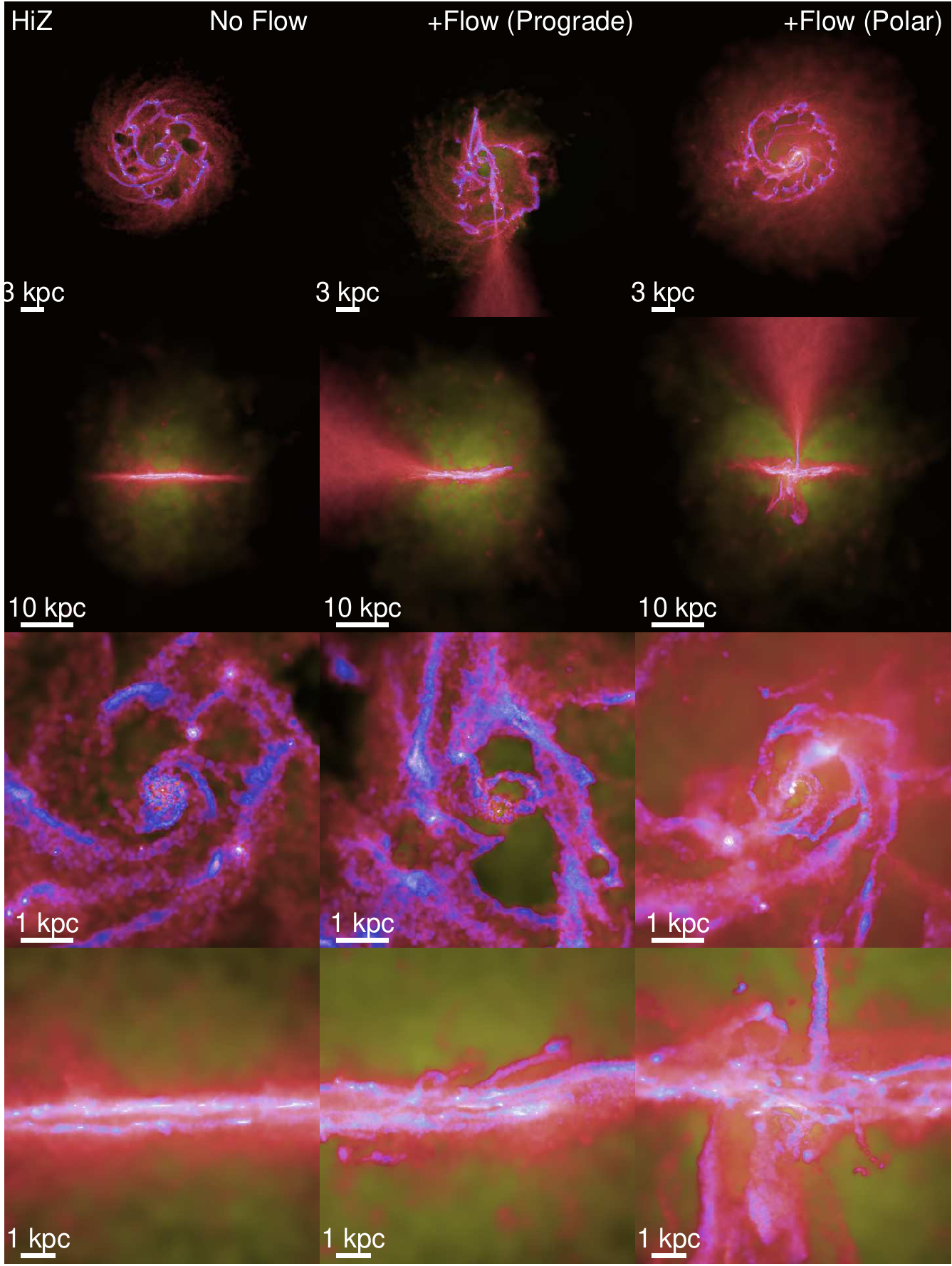}{0.97}
    \caption{Morphology of the gas in simulations of a massive $z\sim2-4$ starburst disk with $\dot{M}_{\ast}\sim200\,\msun\,{\rm yr^{-1}}$. Images are shown at $\sim3.5$ orbital times when the disk is in a feedback-regulated steady-state. We show face-on (upper) \&\ edge-on (lower) projections, at large (halo) scales \&\ small (star-forming disk) scales, as labeled. All models shown include stellar feedback. We compare a case with no inflow ({\em left}), with our ``default'' mock cold flow (inflow rate $\dot{M}\sim200\,\msun\,{\rm yr^{-1}}$, aligned with disk rotation; {\em center}), and with a similar flow oriented perpendicular to the disk rotation ({\em right}). The images are at extremely high resolution/depth. Brightness encodes projected gas density (logarithmically scaled with $\approx6\,$dex stretch); color encodes temperature with blue/white being $T\lesssim1000\,$K molecular gas, pink $\sim10^{4}-10^{5}$\,K warm ionized gas, and yellow $\gtrsim10^{6}\,$K hot gas. The qualitative morphology, with massive kpc-scale molecular complexes, is similar in each case. Warps and out-of-plane streams make the disk appear thicker with flows. Out-flowing winds are seen in all simulations, but are slightly enhanced when inflows are present. 
    \label{fig:coldflow.morph}}
\end{figure}

\vspace{-0.5cm}
\section{The Simulations}
\label{sec:sims}

The simulation method we use is described in detail in \citet{hopkins:rad.pressure.sf.fb} (\S~2 \&\ Tables~1-3; hereafter \paperone) and \citet{hopkins:fb.ism.prop} (\S~2; hereafter \papertwo). We briefly summarize the most important properties here. The simulations were performed with a heavily modified version of the parallel TreeSPH code {\small GADGET-3} \citep{springel:gadget}, in a fully conservative formulation \citep{springel:entropy} which is also density-independent (the ``pressure-entropy'' formulation of SPH) in a manner that allows contact discontinuities and improved fluid mixing \citep{hopkins:lagrangian.pressure.sph}. The artificial viscosity, adaptive timestepping, and smoothing kernel are updated as described therein.\footnote{There has been considerable discussion in the literature regarding subtle numerical effects on inflow properties and the treatment of sub-sonic multiphase flows (see references in \S~\ref{sec:intro}). Our method is specifically designed to better treat this regime. However, we have also re-run our HiZ model with feedback with the alternative ``classical SPH'' formulation in \citet{springel:entropy} and that in \citet{abel:2011.sph.pressure.gradient.est}, as well as varied artificial viscosity and resolution, and find our major conclusions are robust. We believe this is because we focus on the galactic disk, not flow formation and interaction with the hot halo. In the former regime, the shocks and relevant turbulent cascade are highly super-sonic and external forcing (from gravity and feedback) is important; this is the regime where the various numerical approaches agree well \citep[see][]{kitsionas:2009.grid.sph.compare.turbulence,price:2010.grid.sph.compare.turbulence,bauer:2011.sph.vs.arepo.shocks}.} They include stars, dark matter, and gas, with cooling, shocks, star formation, and stellar feedback. 

In our simulations gas follows an atomic cooling curve with additional fine-structure cooling to $10$\,K, allowing for the photo-ionizing background computed in \citet{faucher:ion.background.evol} and gas self-shielding. Metal-line cooling is followed species-by-species for 11 tracked species as in \citet{wiersma:2009.coolingtables,wiersma:2009.enrichment}. The enrichment for each species is followed with the appropriate time dependence with independent yields directly attached to mass and energy return from individual feedback sources described below. Star formation is allowed only in dense, molecular, self-gravitating regions above $n>1000\,{\rm cm^{-3}}$. We follow \citet{krumholz:2011.molecular.prescription} to calculate the molecular fraction $f_{\rm H_{2}}$ in dense gas as a function of local column density and metallicity, and allow SF only from molecular gas. We also restrict star formation to gas which is locally self-gravitating, i.e.\ has $\alpha\equiv \delta v^{2}\,\delta r/G\,m_{\rm gas}(<\delta r) < 1$ on the smallest available scale ($\delta r$ being our force softening or smoothing length). This forms stars at a rate $\dot{\rho}_{\ast}=\rho_{\rm mol}/t_{\rm ff}$ (i.e.\ $100\%$ efficiency per free-fall time); however the galaxy-wide efficiency is generally much lower because of feedback. In \paperone\ and \papertwo\ we show the galaxy structure and SFR are basically independent of the small-scale SF law, density threshold (provided it is high), and treatment of molecular chemistry, because the star formation (and hence galactic structure) is feedback-regulated.

Stellar feedback is included, from a variety of mechanisms:

(1) {\bf Local Momentum Flux} from radiation pressure, 
supernovae, \&\ stellar winds: Gas within a GMC (identified 
with an on-the-fly friends-of-friends algorithm) receives a direct 
momentum flux from the stars in that cluster/clump. 
The momentum flux is $\dot{P}=\dot{P}_{\rm SNe}+\dot{P}_{\rm w}+\dot{P}_{\rm rad}$, 
where the separate terms represent the direct momentum flux of 
SNe ejecta, stellar winds, and radiation pressure. 
The first two are directly tabulated for a single SSP as a function of age 
and metallicity $Z$ and the flux is directed away from the stellar center. 
Because this is interior to clouds, the systems are always optically thick, so the 
latter is approximately $\dot{P}_{\rm rad}\approx (1+\tau_{\rm IR})\,L_{\rm incident}/c$, 
where $1+\tau_{\rm IR} = 1+\Sigma_{\rm gas}\,\kappa_{\rm IR}$ accounts 
for the absorption of the initial UV/optical flux and multiple scatterings of the 
IR flux if the region is optically thick in the IR (with $\Sigma_{\rm gas}$ calculated 
for each particle). 

(2) {\bf Supernova Shock-Heating}: Gas shocked by 
supernovae can be heated to high temperatures. 
We tabulate the SNe Type-I and Type-II rates from 
\citet{mannucci:2006.snIa.rates} and STARBURST99, respectively, as a function of age and 
metallicity for all star particles and stochastically determine at 
each timestep if a SNe occurs. If so, the appropriate mechanical luminosity is 
injected as thermal energy in the gas within the nearest $\sim32$ neighbors of the star particle.

(3) {\bf Gas Recycling and Shock-Heating in Stellar Winds:} Gas mass is returned 
to the ISM from stellar evolution, at a rate tabulated from SNe and stellar mass 
loss (integrated fraction $\approx0.3$). The SNe heating is described above. Similarly, stellar winds 
are assumed to shock locally and inject the appropriate tabulated mechanical 
luminosity $L(t,\,Z)$ as a function of age and metallicity into the gas within a smoothing length. 

(4) {\bf Photo-Ionization \&\ Photo-Electric Heating}: We also tabulate the rate of production of ionizing photons for each star particle; moving radially outwards from the star, we then ionize each neutral gas particle (using its density and state to determine the necessary photon number) until the photon budget is exhausted. Ionized gas is maintained at a minimum $\sim10^{4}\,$K until it falls outside an HII region. Photo-electric heating is followed in a similar manner using the heating rates from \citet{wolfire:1995.neutral.ism.phases}.

(5) {\bf Long-Range Radiation Pressure:} Photons which escape the local GMC (not 
accounted for in (1)) can be absorbed at larger radii. Knowing the intrinsic SED of each star 
particle, we attenuate integrating the local gas density and gradients to convergence. 
The resulting ``escaped'' SED gives a flux that propagates to large distances, and 
can be treated in the same manner as the gravity tree to give the local net incident flux 
on a gas particle. The local absorption is then calculated integrating over a frequency-dependent 
opacity that scales with metallicity, and the radiation pressure force is imparted. 

Details and numerical tests of these models are discussed in \papertwo. 
All energy, mass, and momentum-injection rates are taken as-is from the stellar 
population models in STARBURST99, assuming a \citet{kroupa:imf} IMF, without any free parameters.
Subtle variations in the implementation do not make significant differences to our conclusions. 

We implement the model in three distinct initial disk models spanning a range of galaxy types. 
Each has a bulge, stellar and gaseous disk, halo, and central BH (although to isolate the 
role of stellar feedback, models for BH growth and feedback are disabled). 
At our standard resolution, each model has $\sim 0.3-1\times10^{8}$ total particles, 
giving particle masses of $500-1000\,\msun$ and $1-5$\,pc smoothing lengths, 
and are run for a few orbital times each. In \papertwo, a couple ultra-high resolution runs for 
convergence tests employ $\sim10^{9}$ particles with sub-pc resolution.
The disk models include: 

(1) SMC: an SMC-like dwarf, with baryonic mass $M_{\rm bar}=8.9\times10^{8}\,\msun$ 
(gas $m_{g}=7.5\times10^{8}\,\msun$, bulge $M_{b}=1.3\times10^{8}\,\msun$, 
the remainder in a stellar disk $m_{d}$) and halo mass $M_{\rm halo}=2\times10^{10}\,\msun$. 
The gas (stellar) scale length is $h_{g}=2.1\,$kpc ($h_{d}=0.7$).

(2) MW: a MW-like galaxy, with 
halo $M_{\rm halo}=1.5\times10^{12}$, 
and baryonic 
$(M_{\rm bar},\,m_{b},\,m_{d},\,m_{g})=(7.1,\,1.5,\,4.7,\,0.9)\times10^{10}\,\msun$ 
with scale-lengths 
$(h_{d},\,h_{g})=(3.0,\,6.0)\,{\rm kpc}$. 

(3) HiZ: a high-redshift massive starburst disk, typical of massive star-forming galaxies at $z\sim2-4$; 
$M_{\rm halo}=1.4\times10^{12}\,\msun$ (scaled for $z=2$ halos), 
and baryonic 
$(M_{\rm bar},\,m_{b},\,m_{d},\,m_{g})=(17,\,7,\,3,\,7)\times10^{10}\,\msun$ 
with scale-lengths 
$(h_{d},\,h_{g})=(1.6,\,3.2)\,{\rm kpc}$. 

We consider each model with and without a ``cold flow.'' Motivated by the cosmological simulations cited in \S~\ref{sec:intro}, we initialize the flow as a narrow cone of gas with opening angle $\phi$, infalling at the free-fall velocity (from infinity). We typically assign the flow a small initial turbulent dispersion, $\approx20\%$ of the mean velocity, and a sub-virial initial $T\approx10^{5}\,$K; we have varied these parameters, and found little effect on the results. The flow is initialized so the mean $\dot{M}$ through all conical annuli is constant, towards an impact parameter at a radius $b$ in the disk, with a relative inclination $\theta$. We initialize the flow with sufficient mass and extent to ensure it is maintained for the duration of the simulation. Because such flows are believed to be the ultimate source of the disk gas, our default choice for $b$ is $R_{e,\,{\rm gas}}$ (the effective radius of the gas disk), with a prograde orientation (i.e.\ with the flow having the same sign of angular momentum as the disk), with a narrow ($20\degree$) opening angle, and $\dot{M}$ approximately equal to the maximal halo baryon accretion rate (halo growth rate, for a mean halo of the given mass and redshift, times the universal baryon fraction; $=0.15,\,10.0,\,200\,\msun\,{\rm yr^{-1}}$ for the SMC, MW, HiZ models, respectively). As explained in \S~\ref{sec:intro}, these rates are purposely set to be maximally optimistic; typical (median) halos will have lower accretion rates, and not all of the infall will survive all the way to the galaxy as part of a coherent structure.

\begin{figure}
    \centering
    \plotonesize{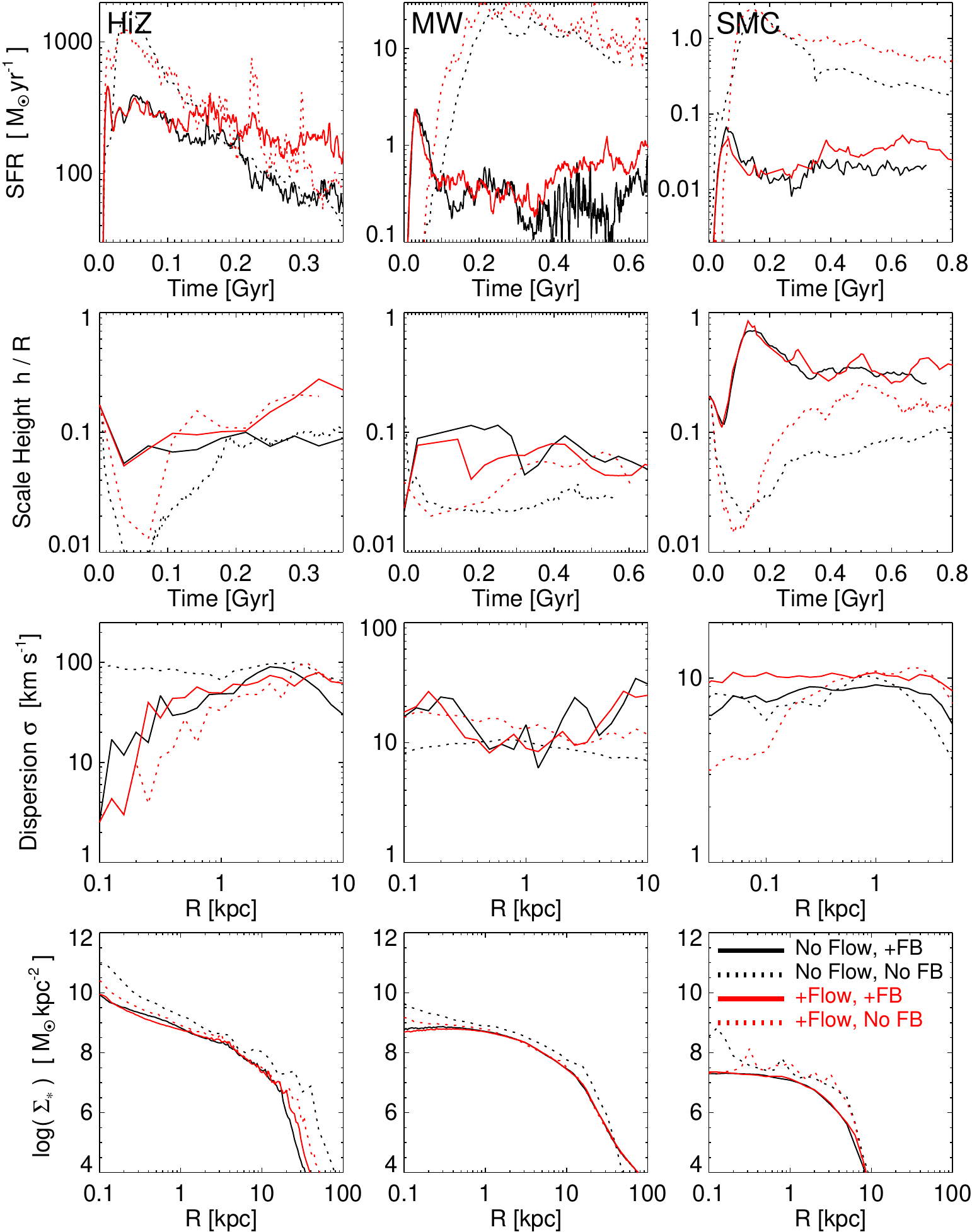}{1.0}
    \caption{Effects of cold ``accretion flows'' on galaxy properties. Each column shows a different galaxy type (\S~\ref{sec:sims}). For each, we show models with/without a flow (using the ``default'' flow parameters) and with/without stellar feedback.
    {\em Top:} SFR vs.\ time. With feedback, models self-regulate near observed SFRs; late-time SFRs with flows are higher because gas is replenished, but both lie on the same Kennicutt-Schmidt relation. Without feedback, both experience runaway local collapse/SF. 
    {\em Second:} Vertical disk scale height $h/R$. Feedback maintains higher values by powering turbulence and isotropizing in-plane gas motions. The flows may help isotropise the gas (especially without feedback) and boost $h/R$ by depositing low-binding energy material above/below the disk.
    {\em Third:} Gas velocity dispersion vs.\ radius (time-averaged). All models maintain $Q\sim1$, giving similar $\sigma$, independent of feedback and flows.$^{\ref{foot:highz.nofb.dispersion}}$
     {\em Bottom:} Final stellar mass profile (at the end of the simulation). Disk instabilities drive some bulge growth in each case, sensitive to feedback, but without a clear dependence on flows.
    \label{fig:coldflow.structure}}
\end{figure}

\begin{figure}
    \centering
    \plotonesize{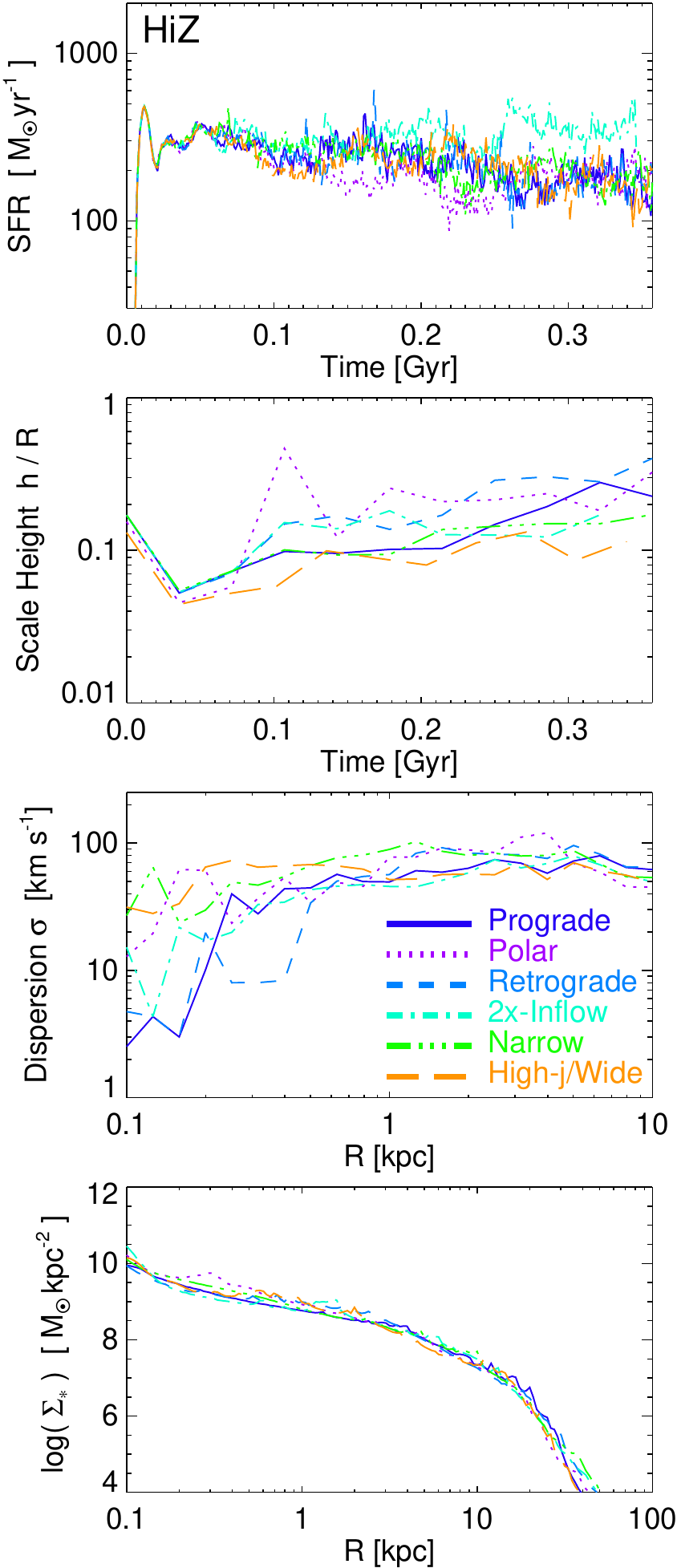}{0.95}
    \caption{Galaxy properties as Fig.~\ref{fig:coldflow.structure}. Here, we focus on the HiZ model (which has a more extreme inflow than the SMC or MW model) but vary the inflow properties. All models include ``cold flows.'' We compare our default cold flow model (prograde -- i.e.\ aligned with the disk, with $\dot{M}_{\rm gas}= 200\,\msun\,{\rm yr^{-1}}$ and impact parameter $b=R_{e}$) to models with different relative flow/disk orientations (a model where the flow is polar, i.e.\ perpendicular to the disk rotation, and a model where the flow is retrograde, i.e.\ directly oppose the disk rotation), with different flow rates (prograde, $400\,\msun\,{\rm yr^{-1}}$), and with different impact parameters (``narrow'' with $b=0.5\,R_{e}$ and cross-section smaller by $1/2$, and ``high-$j$/wide'' with $b=2\,R_{e}$ and cross section larger by $2$). There is very little dependence of the disk properties on these choices except, perhaps, just at the location where the flow meets the disk. Flows with higher angular momentum (impact parameter $>R_{e}$) are similar but appear to have weaker effects on scale-height.
    \label{fig:coldflow.structure.hiz}}
\end{figure}

\vspace{-0.5cm}
\section{Results}
\label{sec:results}

Fig.~\ref{fig:coldflow.morph} compares the gas %and stellar\footnote{\label{foot:stellarimages}The stellar luminosity in each band is calculated from each star particle according to the STARBURST99 model given its age, mass, and metallicity (and smoothed over the appropriate kernel). We then attenuate the stars following the method of \citet{hopkins:lifetimes.letter}: we calculate the total dust column (from the simulated gas) along the line-of-sight to each star particle for the chosen viewing angle (assuming a constant dust-to-metals ratio, i.e.\, dust-to-gas equal to the MW value times $Z/Z_{\sun}$), and apply a MW-like extinction and reddening curve \citep[as tabulated in][]{pei92:reddening.curves}.} 
morphologies of our ``default'' HiZ runs with and without ``flows'' present. The results are broadly similar within the effective radii of the gaseous disks. Inflows clearly lead to more gas distributed outside the star-forming disk, both in the form of streams and filaments falling in, as well as in additional wind material. Over several orbits the inflows modify the outer disk, essentially by replacing it with material on new orbits (determined by the flow parameters). The visual differences are less pronounced in the MW and SMC models.

In Figs.~\ref{fig:coldflow.structure}-\ref{fig:coldflow.structure.hiz}, we examine the effects on galaxy properties. Fig.~\ref{fig:coldflow.structure} examines the ``default'' model in each case, with and without flows present, and with and without feedback. In Fig.~\ref{fig:coldflow.structure.hiz}, we vary the properties of the flow, in our HiZ model with feedback present. We have varied the properties in similar fashion in the MW and SMC runs, but find smaller variation in those cases. This should not be surprising; the gaseous disks are much less strongly self-gravitating in the MW and SMC runs (so less sensitive to external perturbations), and with a much higher inflow rate and SFR the gas ``replenishment'' time in the HiZ case is nearly an order of magnitude faster than in the MW and SMC cases, hence the HiZ disk is ``more dominated'' by inflow and that inflow is much more energetic.

In all cases, as shown in \paperone\ \&\ \papertwo, without feedback, the gas radiates away its support and locally collapses into dense clumps, globally turning most of its mass into stars in a single dynamical time. The SFRs correspond to $\dot{M}_{\ast}\sim M_{\rm gas}/t_{\rm dyn}$, an order-of-magnitude or more in excess of the observed Kennicutt relation. The absolute SFRs decline at later times only because the gas is exhausted, but always exceed the rate given by the Kennicutt relation. The presence of an accretion flow makes no difference to the star formation rate relative to Kennicutt; it simply provides more fuel to sustain the runaway SFR longer. With feedback present, the SFRs self-regulate at a level in good agreement with the observed Kennicutt relation (see \paperone). When feedback is included, the inflow has no effect on the SFR, scaled to Kennicutt; at late times the flows maintain the disk gas densities, slowing the decline in star formation rate.

We next examine the internal gas disk properties. The three-dimensional gas velocity dispersion is plotted (as a function of radius) in the third row of Fig.~\ref{fig:coldflow.structure}. We take the average over all snapshots after the first disk orbital period, when the system is in equilibrium. In \papertwo\ we showed that disks always self-regulate at $Q\sim1$, so the dispersion $\sigma$ is similar regardless of the presence/absence or details of feedback. Without feedback, there are some subtle differences: flows do increase the dispersion in the SMC and MW cases  by a small amount, but actually have the opposite effect in the HiZ case.\footnote{\label{foot:highz.nofb.dispersion}The dispersion in the HiZ case with no feedback or inflow is very large especially at $<1\,$kpc. This is because the extremely rapid star formation (in the absence of feedback), without continuous inflow, rapidly exhausts the central gas, and the dispersion is dominated by hot (volume-filling) gas and the occasional dense gas clump inspiraling near the circular velocity.} With feedback present, these effects are washed out, and we see no statistically significant or meaningful difference with or without inflows. 

The second panel of Fig.~\ref{fig:coldflow.structure} shows the disk scale height $h/R$, which is  related specifically to the vertical disk dispersion $\sigma_{z}$. In \papertwo, we showed that $h$ is different in cases with or without feedback.  Simulations lacking feedback and inflow maintain $\sigma$ via global gravitational instabilities and ``one-way'' collapse, as, e.g.\ a collapsing clump increases in $\sigma$ as it collapses to a resolution-limited small radius. In a disk this gravitationally induced velocity dispersion is preferentially in-plane and leads to large velocity anisotropy, and hence smaller $h/R$ than seen in observations. In the simulations including inflows, the difference between feedback and no-feedback runs is reduced somewhat, i.e., the red dotted line recovers to the solid red line within a couple orbits (though in the lower-mass SMC case, feedback effects on the dispersion are larger so there is still a difference between the results). Visually (Fig.~\ref{fig:coldflow.morph}), the disks appear ``thicker'' with inflows because the streams directly contribute some gas in out-of-plane orbits, and mis-alignment of disk and stream leads over a few orbits to warps; the flows may also help isotropise the gas dispersions, boosting $h/R$ over what it would be without inflows. Comparing to the run lacking {\em both} feedback and inflows, the difference is significant. However, these inflow-driven effects do not much change $h/R$ compared to the no-inflow run with feedback. And the effect on $h/R$ becomes weaker if we systematically increase the impact parameter (and angular momentum) of the inflow; since the dispersions are similar, this appears to be related to the ability of the disk to relax to equilibrium.

We also show the stellar mass profiles at the conclusion of each simulation, after $\sim0.5\,$Gyr of evolution. In \citet{hopkins:clumpy.disk.evol}, we discuss how feedback alters the efficiency of secular evolution, particularly for the HiZ model, comparing the mass profiles with and without feedback. Without feedback, gas experiences runaway local collapse into dense ``nuggets'' which then have no choice but to sink to the galaxy center. This leads to artificially large/rapid bulge formation. We see here the large difference in the bulge formed in the HiZ model with and without feedback (at small radii in the bottom row, the black dotted line lies well above the other lines). Also, without feedback, the outer disk evolves outward as a result of absorbing the angular momentum from these sinking clumps. It has been argued that inflows lead to rapid bulge build up \citep{dekel:2009.clumpy.disk.evolution.toymodel}. We do not see this in our simulations---in fact if anything we find that inflow, in the absence of feedback, leads to {\em smaller} bulges than in simulations lacking both feedback and inflows. If feedback is included, the presence or absence of an inflow has little or no effect on bulge size.

%When inflows are included the effects are still present, but much weaker for the MW/SMC cases, since these are much more stable galaxies. Although the MW model does form a bar, it has a very low gas fraction so this induces little central star formation. We see no evidence that the ``efficiency'' of secular bulge formation is accelerated by the flows. 

\vspace{-0.5cm}
\section{Discussion}
\label{sec:discussion}

We consider idealized simulations of isolated disk galaxies with and without large ``cold streams'', where the cosmological inflows are represented as simplified streams of cold gas free-falling onto the galaxy. This is a crude approximation to the full cosmological case, but unlike in cosmological simulations, our simulations allow us to  add, remove or modify flows while maintaining otherwise identical galaxy properties, i.e., they allow for controlled numerical experiments. The smaller simulation box sizes used also allow us to reach pc-scale and $\sim100\,$yr resolution and include explicit, spatial and time-resolved models for stellar feedback from SNe (Types I \&\ II), stellar winds (O-star and AGB), photoionization, and radiation pressure. 

We show that in an {\em instantaneous} sense -- i.e.\ for otherwise fixed galaxy properties -- the presence or absence of an inflow has only a small effect on the galaxy. The SFR in either case is regulated by feedback: as shown in \papertwo, with explicit feedback, gas cools and collapses until enough young stars are formed to inject sufficient momentum and regulate against further collapse. As a result, the SFR is set by feedback entirely independent of e.g.\ the small-scale star formation law or the presence of cold stream accretion. In other words, the flows have no direct effect on the SFR. They do affect the SFR over long times, not because they affect the dynamics of the ISM, but because they affect the amount of gas in the disk. Without feedback, local collapse runs away in a dynamical time, giving $\dot{M}_{\ast}\sim M_{\rm gas}/t_{\rm dyn}$ ($\sim50$ times larger than observed). This is not altered with flows present. In the absence of feedback, cold streams are unable to prevent runaway local collapse or ``slow down'' star formation via turbulence. Indeed, as shown in \citet{hopkins:excursion.ism}, any mechanism that drives turbulence only on large scales, but does not break up locally self-gravitating regions, is insufficient to stave off runaway local collapse. 

We also find that the velocity dispersions are weakly sensitive to the flow properties. This is consistent with the result in \papertwo\ that a combination of feedback and gravitational instability always drives the velocity dispersion to values such that the Toomre $Q\sim1$ (with any ``excess energy'' efficiently radiated), such that the velocity dispersion does not change upon adding or removing individual feedback coupling or turbulent driving mechanisms. It follows that the stability properties of the galaxies are not significantly altered by the presence of inflows. 

Nor do inflows  appear to strongly alter the rate of secular evolution or angular momentum transfer in the disk. Disk instabilities, which are present in all the simulations shown here, are primarily sensitive to local properties. The inflow we model is smooth and extended, and so does not introduce a strong resonant response \citep[see e.g.][]{donghia:2010.tidal.resonances}, unlike the case in a galaxy merger. 

Feedback efficiently powers turbulence and isotropises the dispersions, so with feedback on we see little difference in the disk scale-height when in inflow is turned on or off. However, in runs without feedback, the velocity dispersion is largely in-plane in the no-flow case, so adding an inflow ``stirs'' the disk and boosts its thickness significantly. We note, though, that these no-feedback, inflow-stirred disks are still thinner than disks with feedback present. Flows do  make the disks appear thicker,  by contributing some out-of-plane star-forming gas directly, and, over time, generating warps and misalignments. This is not a consequence of the flow dynamically disturbing, stirring, or driving the disk, but rather of the disk material (coming from the flow) having a different angular momentum distribution (with mis-aligned material) in our idealized, isolated galaxy models. With sufficient time (if the flow is stable), the situation will likely  relax into a new disk aligned with the flow, though situations where the flow is constantly changing in orientation (if they occur in cosmological simulations) would be particularly interesting to study in the future. 

If we examine the location where inflow meets disk, we see a mix of behaviors. Some diffuse material undergoes a series of radiative shocks, but the relative velocities in each are sufficiently small that this places material near the peak of the cooling curve and so the energy is efficiently dissipated. More dense streams and clumps fall in ballistically, and can penetrate the disk, forming the mis-aligned streams discussed above, which are gradually torqued into equilibrium orbits. 

We also note that the feedback-driven outflows or winds discussed in detail in \citet{hopkins:stellar.fb.winds}, are not strongly altered with cold flows present, although overall the outflow mass-loading rates ($\dot{M}_{\rm wind}/\dot{M}_{\ast}$) are enhanced by factors $\sim1.5-2$ (reflected in the extended outflows in Fig.~\ref{fig:coldflow.morph}), because the inflow includes material which is marginally bound. This marginally bound inflowing material can be accelerated by radiation pressure and hot gas ram pressure at large radii, i.e., it is accelerated off of the surface of the stream. 

The ``cold flow'' models here represent extreme cosmological situations, in that they have mass inflow rates equal to the entire baryonic halo accretion rate and assume $100\%$ of the infall is in ``cold mode.'' They are infalling at the free-fall velocity, reaching the galaxy, and there is (by design) no hot halo to resist their infall or strip the streams. They also have impact parameters very close to the disk effective radii. More realistic situations in cosmological simulations may produce weaker effects than those here. This is especially the case in lower-redshift or more massive halos like the MW, where much of the gas is probably shocked to virial temperatures with long cooling times. Furthermore, in massive halos, even at $z\sim2$ gas delivered from minor mergers might be an important channel of gas supply \citep{keres:cooling.revised}. Moreover, recent simulations using different numerical methods have questions some early conclusions about the dynamics of infalling gas, arguing for example that the streams may be more smooth and diffuse, and more efficiently stripped/mixed by halo gas as they approach the disk, than some earlier calculations (mostly using ``classical SPH'') had estimated \citep{vogelsberger:2011.arepo.vs.gadget.cosmo,sijacki:2011.gadget.arepo.hydro.tests,keres:2011.arepo.gadget.disk.angmom,bauer:2011.sph.vs.arepo.shocks}.

In many ways, our results confirm the ``conventional wisdom'' in this subject. However, one way in which accretion could introduce stronger perturbations than what we consider here is if it were very ``clumpy,'' with a large fraction of the incoming mass in units with sizeable masses relative to the disk \citep[see particularly][]{dekel:cold.streams}.  \citet{hopkins:disk.heating} discuss the resonant response to radially infalling clumps/galaxies, and \citet{donghia:2010.tidal.resonances} consider non-merging passages; both suggest that significant effects could occur if the clumps have masses $\gtrsim10\%$ of the disk, effectively acting as minor mergers. And indeed there may be a large number of such clumps and/or mergers ``carried in'' with the flows \citep[see e.g.][although also see the numerical caveats above]{keres:hot.halos,keres:cooling.revised,dekel:cold.streams}. The response of disks to minor mergers and harassment, however, is well-developed and studied in a number of other contexts, and is outside the scope of our investigation here. But to the extent that this is important, our results suggest it may be separable into a merger-like contribution from individual massive clumps falling in, and a weaker contribution from the ``smooth accretion flow.''

We are of course {\em not} arguing that ``cold'' accretion is unimportant for galaxy formation. None of our arguments change whether gas enters the halo without undergoing virial shocks or with cooling times short compared to dynamical times and so reaches the galaxy rapidly. This gas is the  supply for star formation on cosmological timescales; therefore it (with the effects of subsequent feedback) determines the disk gas mass, stellar mass, and angular momentum content of the galaxy, which {\em in turn} determine the SFR, velocity dispersion, and secular stability \citep[see e.g.][]{brooks:2009.coldflow.disk.assembly,federrath:2012.sfr.vs.model.turb.boxes}. Both theoretical \citep{keres:hot.halos} and observational \citep[e.g.][]{daddi:2007.sfr.z2.exhaustion.time} arguments indicate that continuous gas accretion is necessary to to explain long term star formation activity of galaxies. For example, on cosmological timescales, the ``equilibrium SFR'' of galaxies will be set in part by the rate of inflow: the gas consumption rate is $\dot{M}_{\ast}+\dot{M}_{\rm wind}$, where $\dot{M}_{\rm wind}\propto \dot{M}_{\ast}$ for otherwise similar properties (\paperthree). If this falls below the infall $\dot{M}_{\rm acc}$, a gas reservoir inevitably builds up, increasing the SFR until $\dot{M}_{\rm acc}\sim \dot{M}_{\ast}+\dot{M}_{\rm wind}$. However, SF (and the generation of outflows) does not directly  ``know'' about the inflow. Rather, inflows (in competition with SF and outflow) determine the gas supply, which determines the SFR locally needed to prevent runaway local collapse, which determines the actual SFR, which then determines the outflow properties.

In short, we argue that the role of accretion is to {\em cosmologically} determine the ``initial'' global properties of the galaxy, e.g., the gas  and angular momentum supply rate. For a given set of those global properties, the local instantaneous properties of the galaxy are set by a combination of local dynamical and feedback processes that operate on much smaller spatial and time scales.\\

\vspace{-0.9cm}
\acknowledgments 
Support for PFH was provided by NASA through Einstein Postdoctoral Fellowship Award Number PF1-120083 issued by the Chandra X-ray Observatory Center, which is operated by the Smithsonian Astrophysical Observatory for and on behalf of NASA under contract NAS8-03060. NM is supported in part by NSERC and by the Canada Research Chairs program. DK acknowledges support from NASA through Hubble Fellowship grant HSTHF-51276.01-A.\\%We thank Todd Thompson for helpful discussions and contributions motivating this work.
%Support for PFH was provided by the Miller Institute for Basic Research 
%in Science, University of California Berkeley.
%EQ is supported in part by NASA grant NNG06GI68G and 
%the David and Lucile Packard Foundation.  
%\\

%\vspace{-0.2cm}
\bibliography{/Users/phopkins/Documents/work/papers/ms}
%\bibliography{ms}

\end{document}